\documentclass{JINST}
\usepackage{subfigure}
\usepackage{paralist}
\newcommand{\refF}[1]{figure~\ref{#1}}

\title{Electron transparency of a Micromegas mesh}
\author{K.~Nikolopoulos$^{a,b}$\thanks{Corresponding author},
P.~Bhattacharya$^c$,
V.~Chernyatin$^b$, and
R.~Veenhof$^d$\\
\llap{$^a$}Physics Department, University of Athens, Athens, GR 15771, Greece\\
\llap{$^b$}Physics Department, Brookhaven National Laboratory, Upton, NY 11973, U.S.A.\\
\llap{$^c$}ANP Division, Saha Institute of Nuclear Physics, Kolkata 700064, India\\
\llap{$^d$}Physics Department, University of Wisconsin, Madison, WI 53706-1481, U.S.A.\\
\\
E-mail: \email{Konstantinos.Nikolopoulos@cern.ch}}

\abstract{Measurements of the electron transparency of a Micromegas mesh are compared to simulations.
The flux conservation argument is shown to lead to inaccurate estimates of the transparency,
the importance of accurate geometric modelling of the mesh is discussed
and the effect of the dipole moment of the mesh is demonstrated.
This study provides a validation of the microscopic simulation methods specifically
developed for micropattern devices where the characteristic dimensions are of the
same order of magnitude as the electron mean free path in the gas.}


\begin{document}
\section{Introduction}
In Micromegas detectors~\cite{Barouch:1998rt,Giomataris:1995fq},
a micro-mesh separates the drift region from the amplification region.
We report on a detailed study, both experimental and theoretical, of the
electron transparency of the mesh.
The transparency enters in the gain calibration of the detector
and is particularly important for a new generation of multi-gap detectors
which are less prone to discharges~\cite{fabien}. 

These measurements are sensitive to electron transport at the micron-scale,
and thus test the Magboltz-based~\cite{magboltz} microscopic tracking algorithms
in Garfield~\cite{heinrich,Veenhof:2009zza} as well as the field calculation methods.

\section{Experimental set-up}
\label{sec:transpSetup}
\begin{figure}[ht]
\centering
\includegraphics[height=0.425\linewidth]{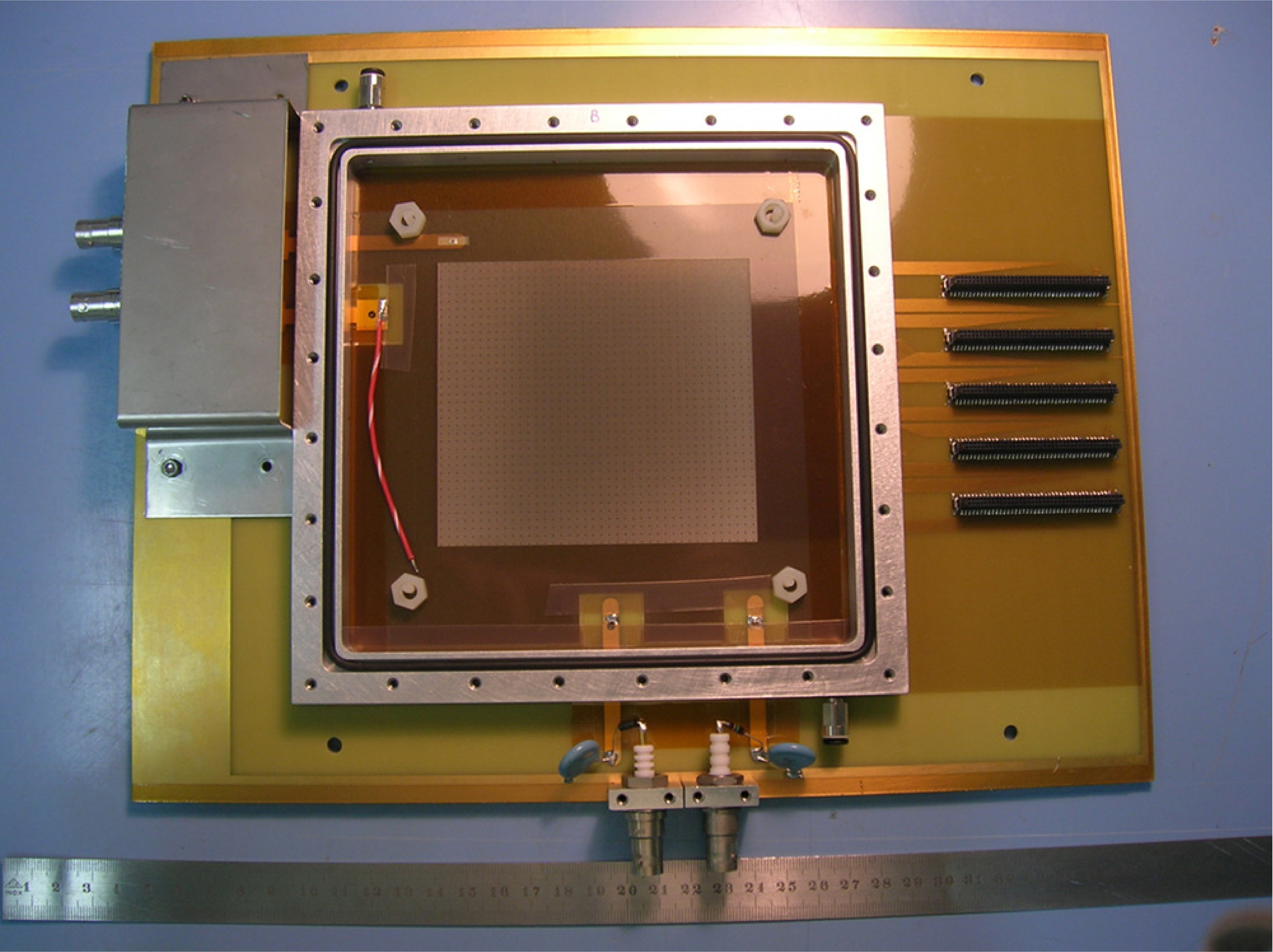}
\includegraphics[height=0.425\linewidth]{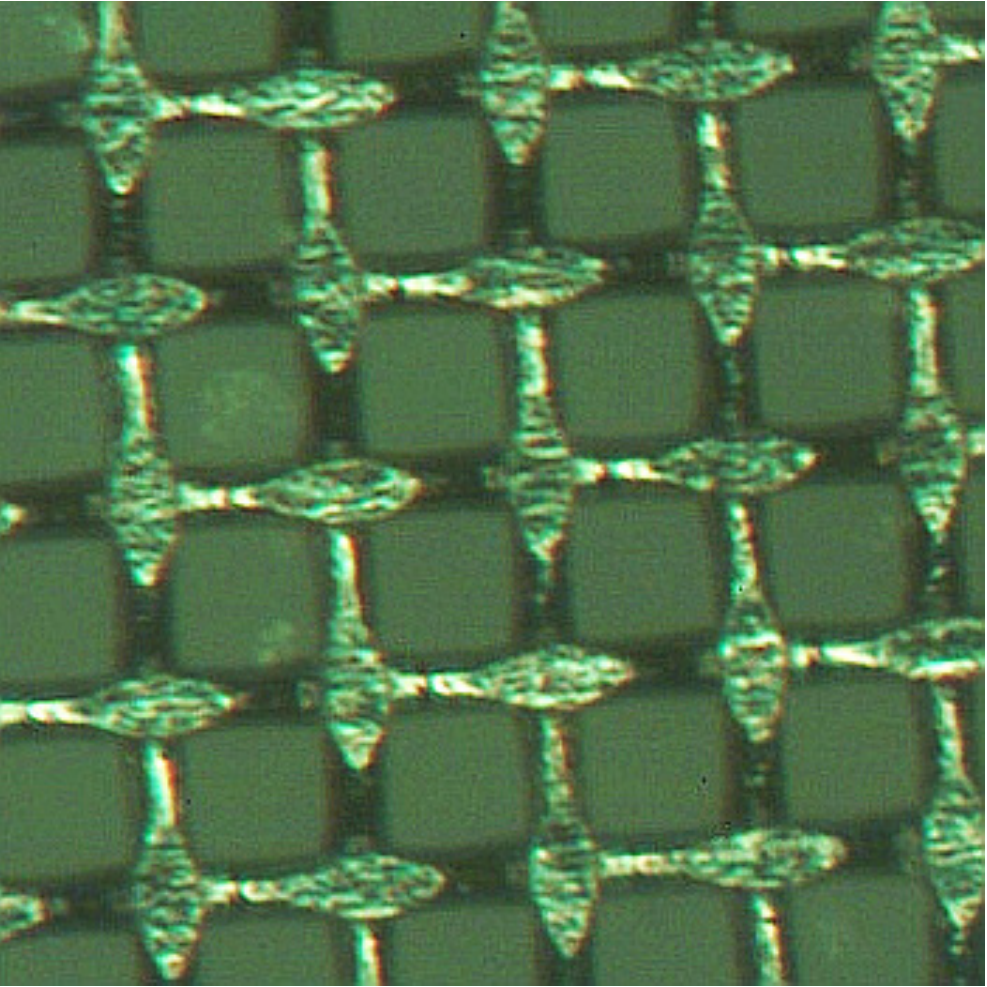}
\caption{Left: the Micromegas chamber used for the measurements.
Right: close-up of the calendered mesh;
note the flattening of the wires at the intersections.
Photo: Rui de Oliveira.}
\label{fig:micromegasandmesh}
\end{figure}
We have measured the transparency of a $100~\mathrm{mm} \times 100~\mathrm{mm}$
bulk-Micromegas~\cite{Giomataris:2004aa} produced by CERN/EN-ICE,
equipped with a calendered woven micro-mesh
similar to the Micromegas detector of the TPC of the T2K experiment~\cite{Giganti:2009zz},
see \refF{fig:micromegasandmesh}.
The stainless-steel wires had a diameter of $18~\mu\mathrm{m}$,
a pitch of $63.5~ \mu\mathrm{m}$ and
the thickness of the mesh at the intersections of the wires was $30~\mu\mathrm{m}$~\cite{ruiprivate}.
The amplification gap thickness was $128~\mu\mathrm{m}$ and the drift distance $2~\mathrm{mm}$.
The chamber was flushed with $\mathrm{Ar} ~ 85~\% ~ \mathrm{CO}_2 ~ 15~\%$.

We have used an $^{55}\mathrm{Fe}$ source, taking data at a rate of $200~\mathrm{Hz}$.
The amplification field was $41.4 ~\mathrm{kV/cm}$ which gave a gas gain of $3\cdot 10^3$.
The electric field in the drift region $E_\mathrm{drift}$ ranged from $110~\mathrm{V/cm}$ to $6.5~\mathrm{kV/cm}$.
Signal amplitudes were calculated from the position of the photo peak,
which has a resolution of $11 ~\%$.
We estimate the mesh transparency as the ratio of the signal amplitude at a given drift field
over the mean signal amplitude for $E_\mathrm{drift} < 500~\mathrm{V/cm}$.
The typical uncertainty of the transparency is $1~\%$.
The drift and amplification field are defined in this paper as the ratio of voltage difference
over distance, respectively between cathode and mesh and between mesh and anode.

The anode of the detector was segmented in strips of $250~\mu\mathrm{m}$ width.
The output of the 76~strips nearest to the source was summed and
passed through a charge sensitive pre-amplifier acting as an integrator.
Its decay time constant was $250~ \mu\mathrm{s}$, three orders of magnitude larger
than the $200~\mathrm{ns}$ signal duration.
Subsequently, it was fed to a unipolar semi-Gauss shaping amplifier (ORTEC model 571)
with an integration time of $1 ~\mu\mathrm{s}$.
Finally, the measurements where digitised and recorded through a National Instruments PC-interface card.

\section{Simulation}
\begin{figure}[ht]
\centering
\includegraphics[height=0.485\textwidth]{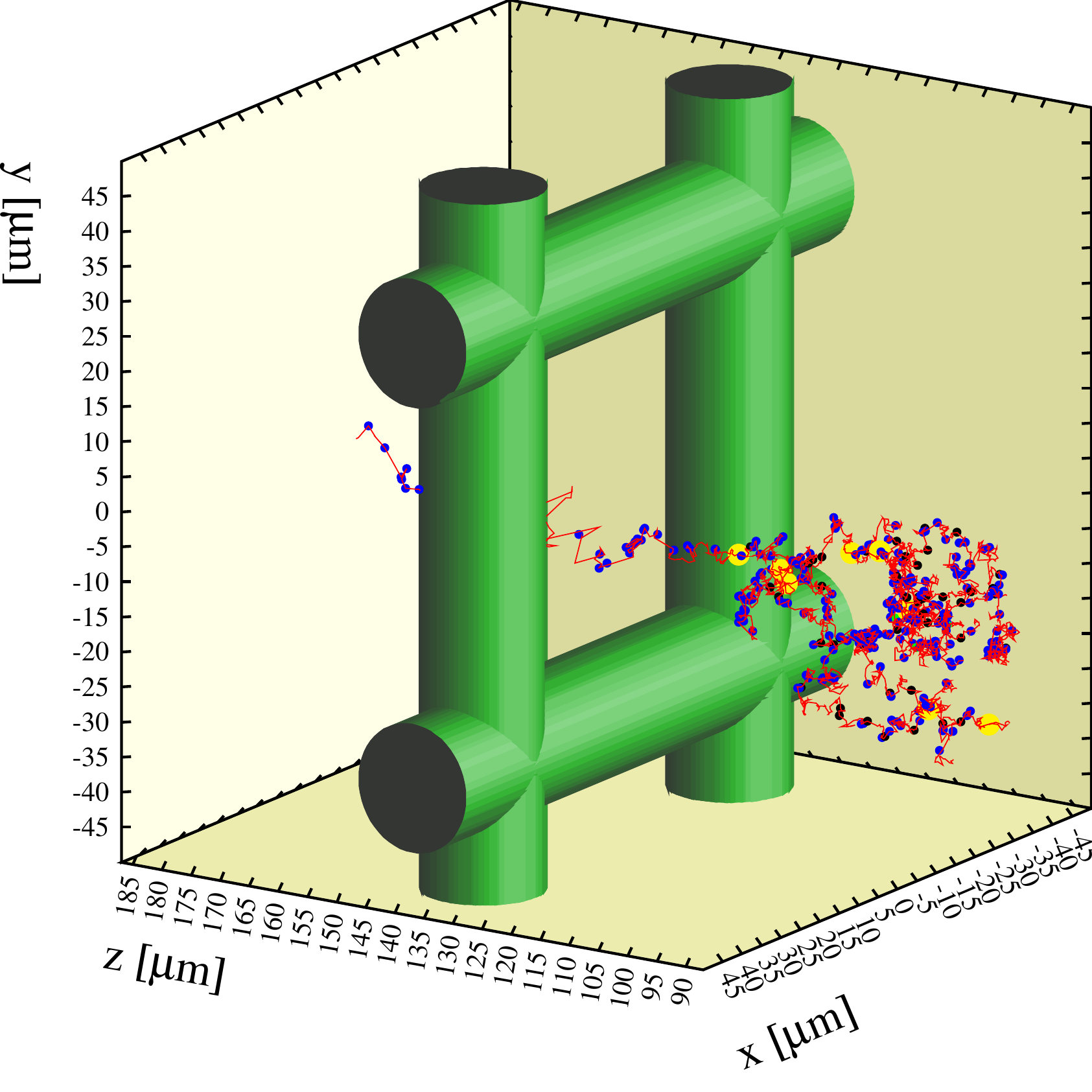}
\hspace{0.01\textwidth}
\includegraphics[height=0.485\textwidth]{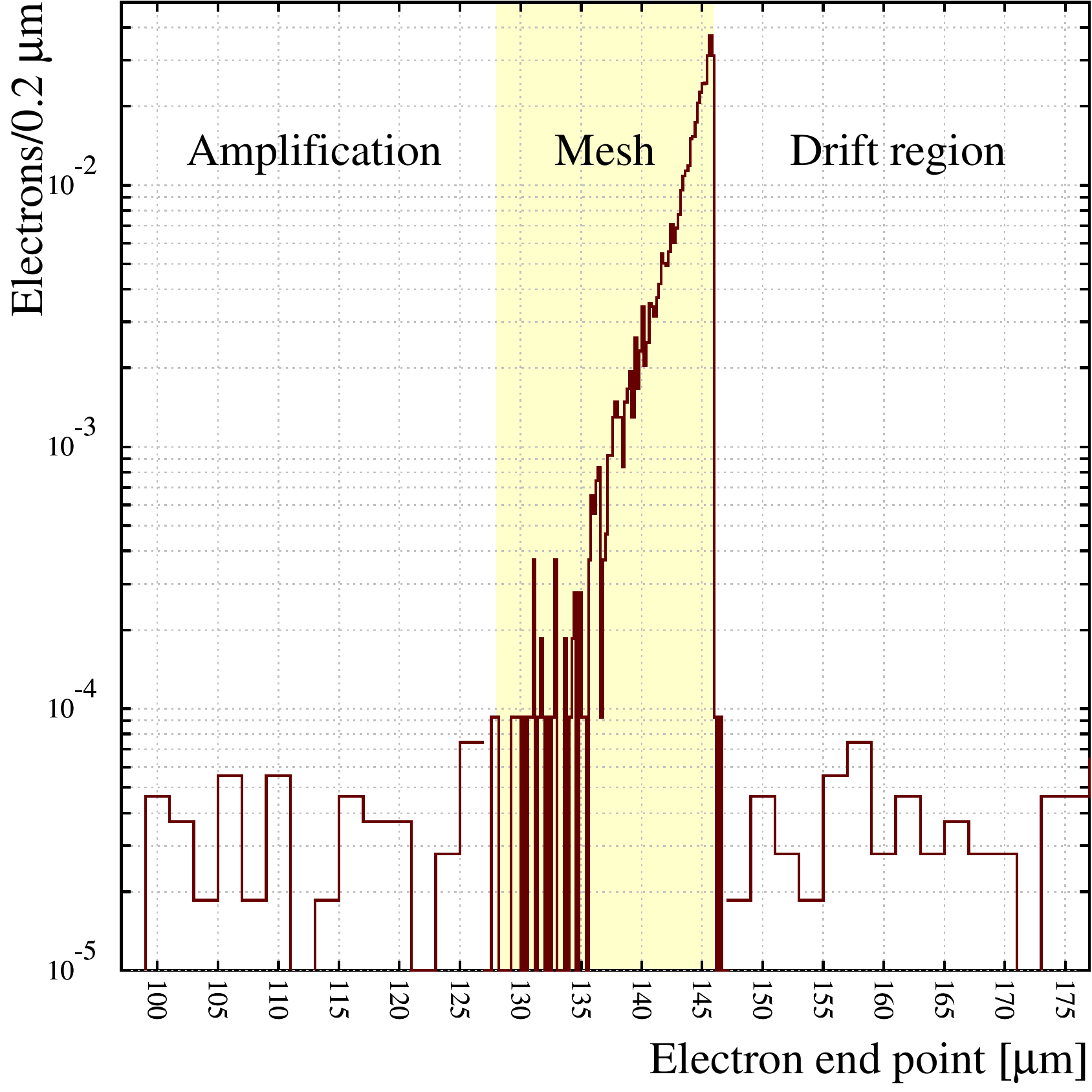}
\caption{Left: the basic mesh cell of the Micromegas,
constructed with cylindric wires.
The figure also illustrates microscopic electron tracking:
every other electron collision is connected with red lines,
blue dots mark inelastic collisions with $\mathrm{CO}_2$
in which vibrational, rotational and polyad states are formed,
black dots show $\mathrm{Ar}^{*} 3p^5 4s, ~ 3p^5 4p ~ ...$ excitations and
large yellow dots represent ionisations.
Note that the avalanche starts after the mesh.
Also well visible is the decrease in mean free path when going from the drift to the amplification zone.
Right: Electrons are predominantly lost on the drift side of the mesh,
irrespective of the transparency of the mesh.
This is shown here by summing histograms for voltage conditions
ranging from transparency to intransparency.
The losses in the gas volumes are due to electron attachment by $\mathrm{CO}_2$.
}
\label{fig:systematics}
\end{figure}

\label{sec:simulation}
\label{sec:transimulation}
Since no exact solution is known for the field around a mesh,
we have approximated the field using the finite element method and using the boundary element method.
Despite the similarity in name, these techniques have strikingly different properties.
The finite element method meshes the volume and parametrises the potential locally by means
of quadratic polynomials.
This potential is not as a rule a solution of the Maxwell equations and
the resulting fields can be discontinuous across element boundaries,
but the potential boundary conditions are strictly respected.
The boundary element method meshes the surfaces of the electrodes and dielectrics.
The fields are Maxwell-compliant and continuous,
but the voltage boundary conditions are not necessarily strictly respected.

For the finite element approach, we have used the second-order tetrahedral elements with parabolic faces
of Ansys~\cite{ansys}.
These elements are well suited for curved structures such as mesh wires.
Numerically, they are efficient since the iterative calculation of the isoparametric coordinates
converges rapidly while the Jacobian of the final iteration serves to compute the potential gradient.
The boundary element calculations were performed using neBEM~\cite{nebem2006, nebem2009}
which offers two elements with which a mesh can be modelled:
one-dimensional thin-wire segments and
three-dimensional polygonal approximations of cylinders.
Wire elements are computationally attractive but
the field is calculated using the thin-wire approximation.
As a result, the voltage at one wire radius from the axis will
only on average equal the imposed surface potential.
When cylindric elements are used, the voltage boundary condition is applied
to each of the surface panels of the cylinder.

Conservation of the flux of the electric field combined with the assumption that
the electric flux is proportional with the electron flow,
leads to a mesh transparency of $\phi_\mathrm{A}/(\phi_\mathrm{A}+\phi_\mathrm{M})$ where 
$\phi_\mathrm{A}$ is the electric flux from the cathode to the anode and
$\phi_\mathrm{M}$ the flux from the cathode to the mesh.
The flux ratio is easily calculated by means of Runge-Kutta-Fehlberg integration of the
drift velocity vector.

The flux argument assumes that electrons actually follow the electric field, and thus neglects diffusion.
To assess the impact of diffusion we have also calculated the transparency using 
microscopic electron tracking~\cite{heinrich,Veenhof:2009zza}.
This technique tries to reproduce electron transport at the molecular level.
Electrons follow a vacuum trajectory between collisions.
At each collision, one of the processes available in Magboltz~\cite{magboltz} is selected at random,
weighed by the probabilities for the various processes to occur, considering the electron energy prior to impact.
An electron can lose energy in inelastic collisions,
it may be lost to attachment, excite an atom or cause ionisation.
Penning transfer, which has a comparatively minor impact 
in $\mathrm{Ar}\mbox{-}\mathrm{CO}_2$ mixtures~\cite{ozkan},
was not taken into account.

For both integration techniques, the transparency was calculated
by drifting electrons from randomly distributed points in the drift region,
$100 ~ \mu\mathrm{m}$ from the mesh.
At this distance, the flux is uniform within the numerical precision and diffusion ensures that
the transparency is decorrelated from the starting point.
The transparency is estimated as the fraction of electrons arriving
in the amplification region.

We do not present results for the more time consuming microscopic avalanche technique 
because it yielded essentially the same results as microscopic electron tracking in a few
test cases.
This is not surprising given that electron losses are concentrated on the drift side of the mesh
while the electrons start multiplying only after they have passed the mesh,
as seen in figure~\ref{fig:systematics}.

\section{Results}
\label{sec:measurementAndComparison}
\begin{figure}[!t]
\centering
\includegraphics[height=0.42\textwidth]{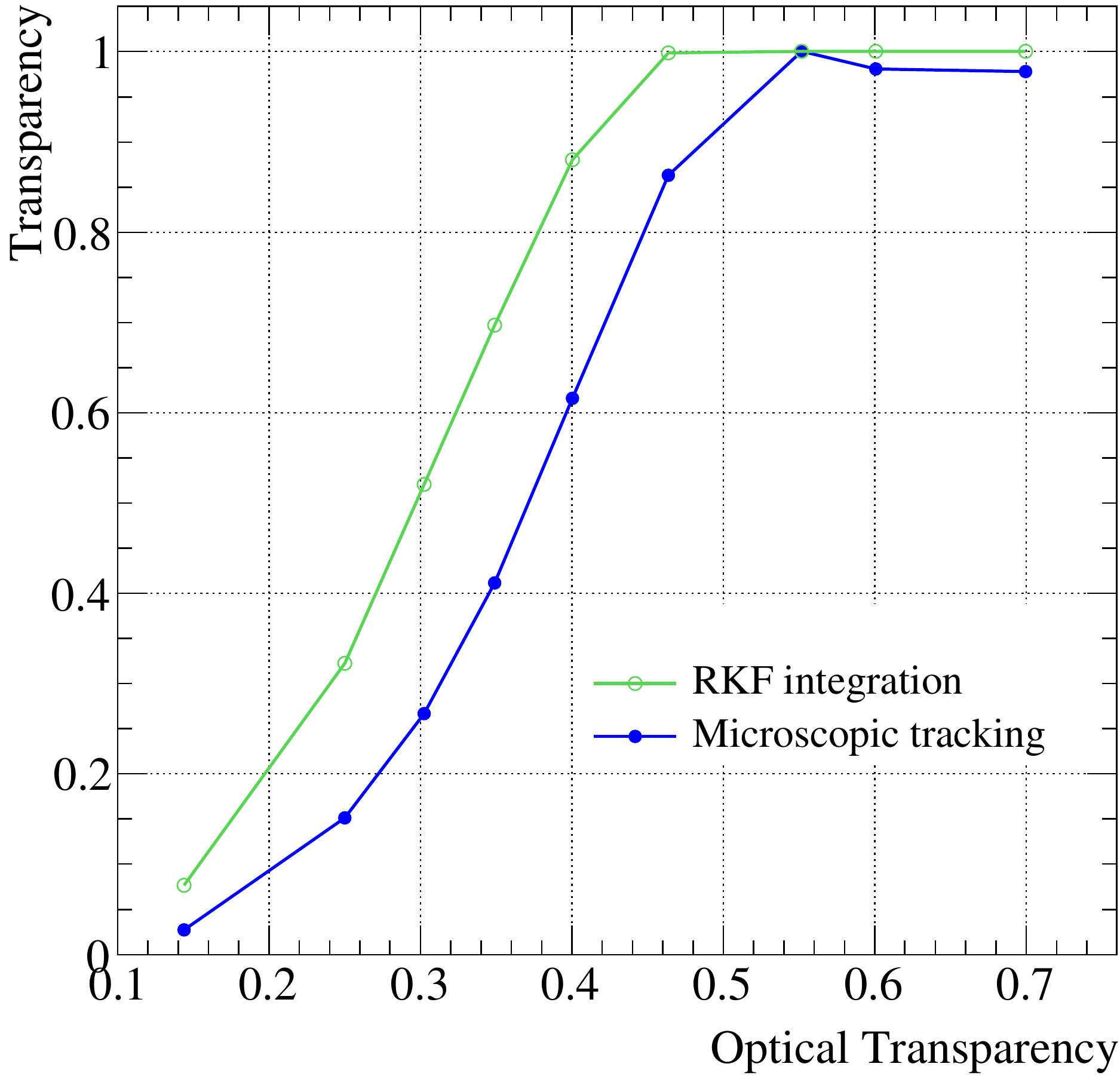}
\caption{Calculated electron transparency as function of the optical transparency of the mesh.
The optical transparency is defined as the fraction of the perpendicular projection
of the mesh not obscured by wires.
The diameter of the (cylindric) wires is $18~\mu\mathrm{m}$ throughout and the wire pitch varies.
The amplification field is $41.4 ~\mathrm{kV/cm}$ and the drift field $1 ~\mathrm{kV/cm}$.\label{fig:optical}}
\vspace{0.2cm}
\includegraphics[height=0.42\textwidth]{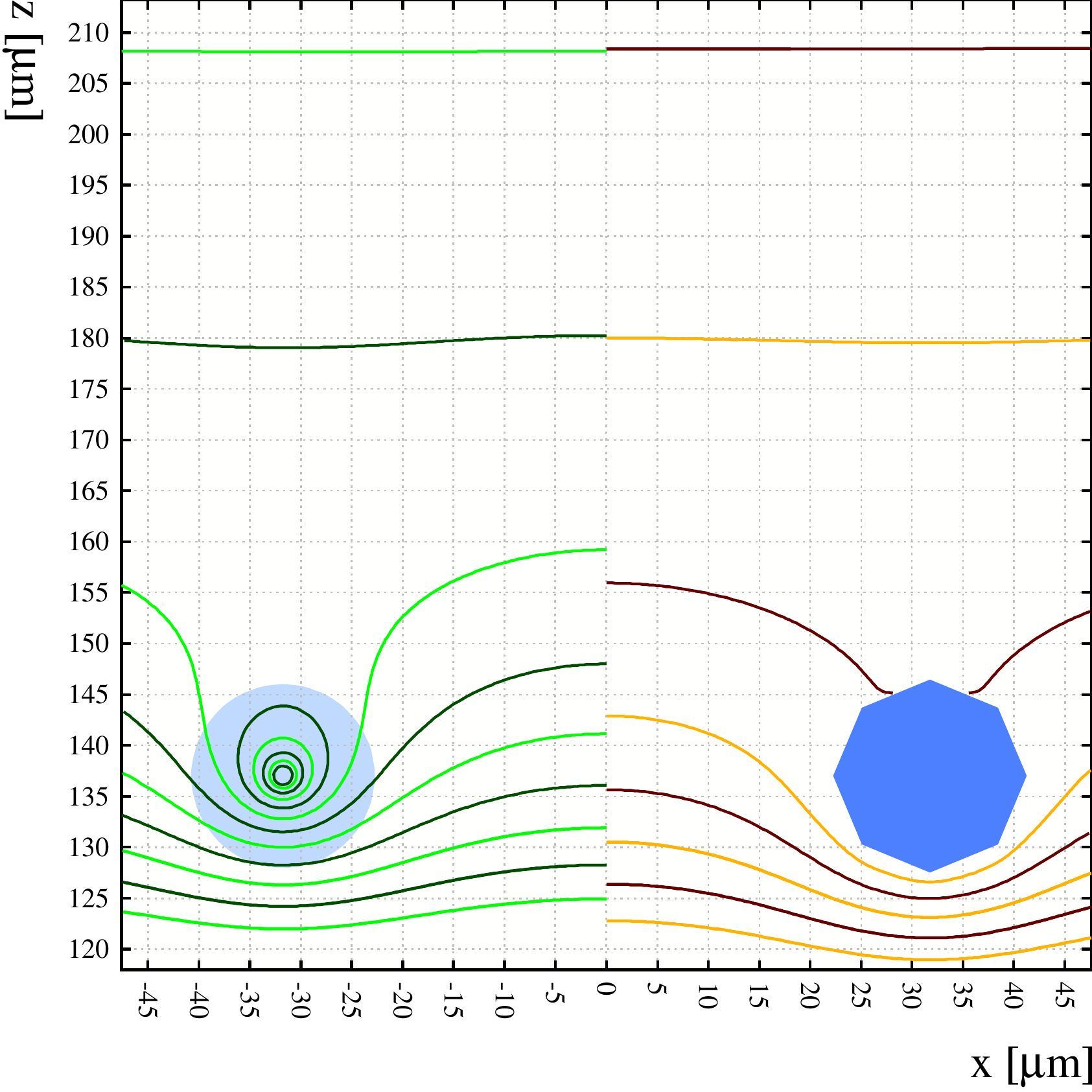}
\caption{Equipotentials in the mid-plane of a mesh-hole,
in the thin-wire approximation of the mesh (green) and
using octagonal approximations of solid cylinders (brown).
The sixth contour from the bottom corresponds with the nominal mesh potential.
At the drift field of $3.3~\mathrm{kV/cm}$ for which the plot was made,
the transparency computed with wire elements is $15~\%$ larger than the data,
see figure~\protect\ref{fig:result}.
The amplification field is $41.4 ~\mathrm{kV/cm}$.\label{fig:contours}}
\end{figure}

\begin{figure}[t]
\includegraphics[width=0.49\textwidth]{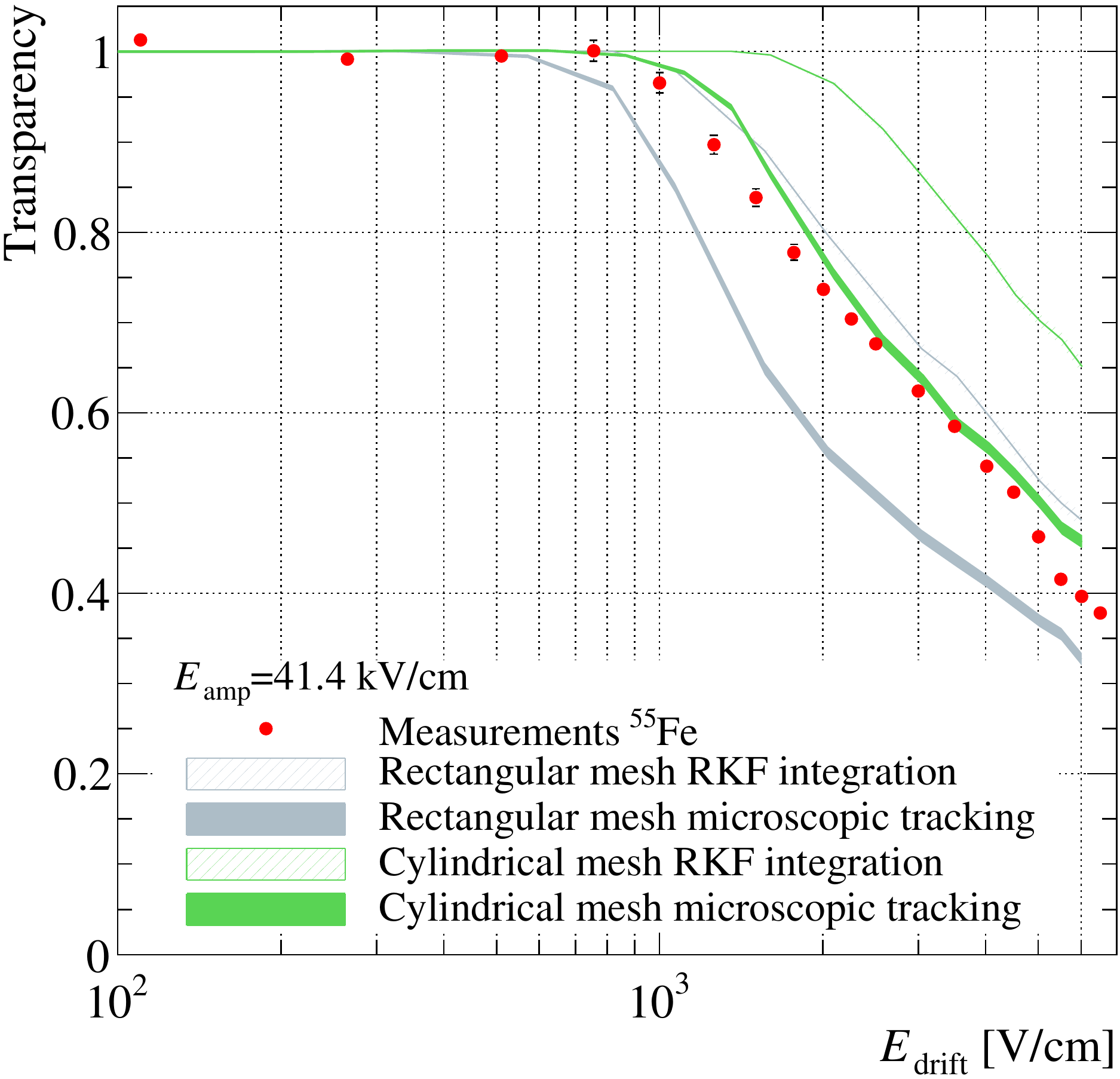}
\hspace{0.01\textwidth}
\includegraphics[width=0.49\textwidth]{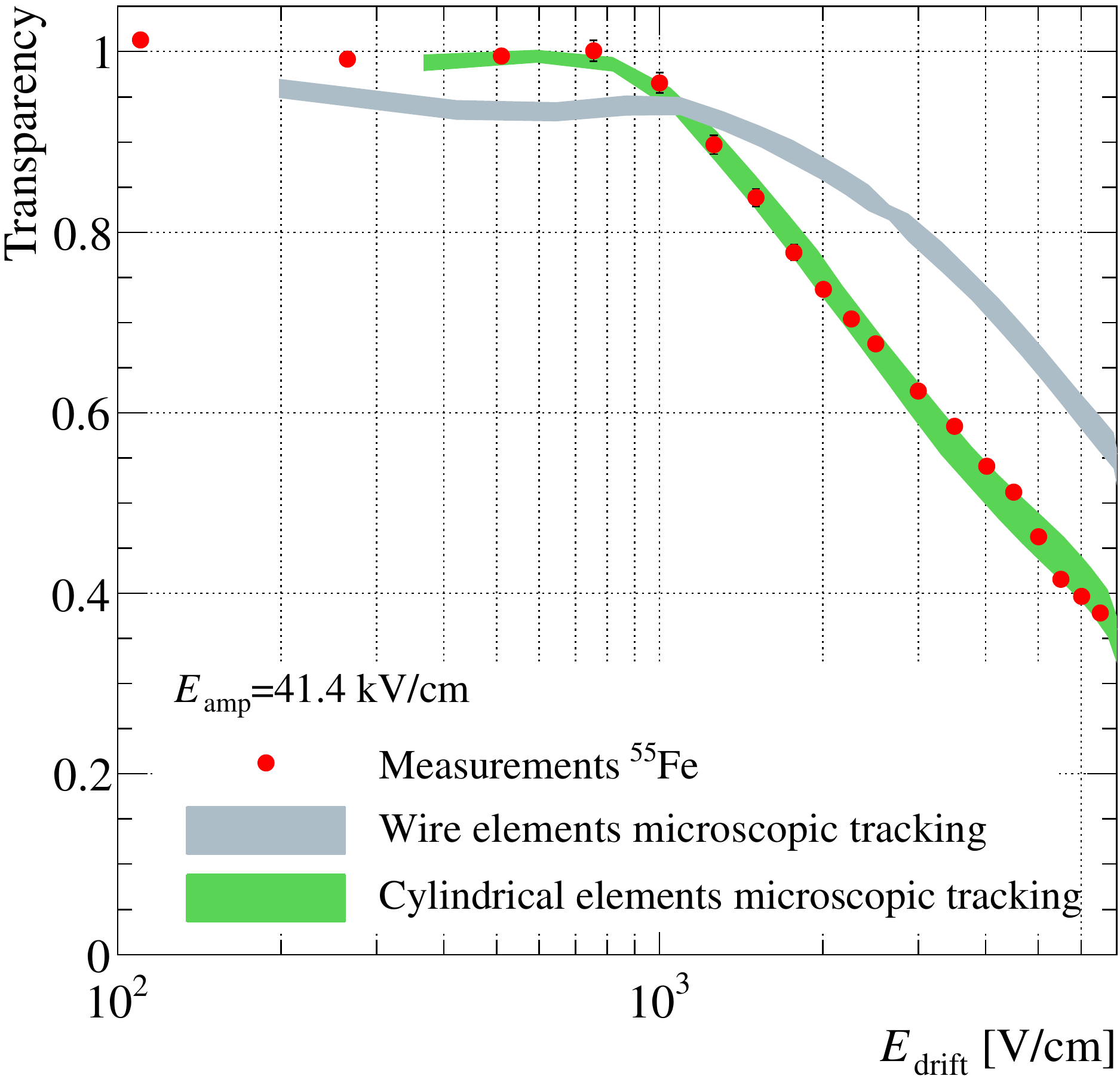}
\caption{Left: measured relative amplitude for the $5.9~\mathrm{keV}$ line of $^{55}\mathrm{Fe}$
as a function of $E_\mathrm{drift}$ for a constant amplification field.
They are compared with transparency estimates obtained with microscopic electron
tracking and with zero-diffusion integration, for both cylindric and rectangular wires.
Right: the same measurements compared with boundary element calculations 
using wire elements and using solid cylindric elements.
The data has been normalised in the region $E_\mathrm{drift} < 600 ~ \mathrm{V/cm}$.
Similarly, the calculations have been scaled up by  $1.4~\%$.
The width of the curves reflects the statistical accuracy of the calculations.}
\label{fig:result}
\end{figure}

The measurements and the simulations are compared in \refF{fig:result}.
The measured transparency is constant for $E_\mathrm{drift} < 800~\mathrm{V}/\mathrm{cm}$,
beyond which electrons increasingly often hit the mesh.
The transparency decreases to reach 40~\% at $E_\mathrm{drift} = 6 ~\mathrm{kV/cm}$.
The microscopic transparency calculation agrees to 5~\% with
the measurements over the entire drift field range.
This is true both for finite element and for boundary element field calculations.
In contrast, the zero-diffusion transparency decreases
only from $E_\mathrm{drift} = 1.7~ \mathrm{kV}/\mathrm{cm}$ and the transparency is
overestimated by 20~\% at $E_\mathrm{drift} = 2 ~\mathrm{kV/cm}$.
The difference between microscopic tracking and flux estimates is also clearly visible
when predicting the electron transparency as a function of the wire pitch,
see figure~\ref{fig:optical}.

The sensitivity of the transparency to details of the geometry
can be demonstrated by comparing two shapes of the mesh wires:
\begin{inparaenum}[\itshape a\upshape)]
\item rectangular mesh wires with sides equal to the actual mesh wire diameter,
a simplification that is frequently made in such calculations; and
\item realistic cylindric mesh wires (\refF{fig:systematics}).
\end{inparaenum} 
In both cases, we assume that the mesh wires pass through one-another at the intersections --
the mesh thickness at the wire intersections is in reality 
larger than the wire diameter.
A rectangular wire mesh has the same optical transparency as a cylindric wire mesh,
but it has a lower electron transparency because of the larger volume and larger surface area of the wires.
The microscopic transparency for rectangular wires starts to decrease already at 
$E_\mathrm{drift} = 700 ~ \mathrm{V/cm}$ and the transparency is 20~\% lower than
the measurements at $E_\mathrm{drift} = 2 ~ \mathrm{kV/cm}$.
The zero-diffusion calculation for rectangular wires happens to be in fair agreement with the measurements.
These calculations have been performed using the finite element technique, but could equally well
have been done with the boundary element method.

Simulations also reveal the importance of the dipole moment of the grid.
In normal operation, the mesh as a whole is negatively charged -- it repels electrons.
To ensure an equal potential on both surfaces, there is additional positive charge on the drift side
and an equal amount of negative charge on the amplification side.
This leads to a dipole moment, as illustrated in figure~\ref{fig:contours}.
The thin-wire approximation neglects this dipole moment and
leads to incorrect estimates of the transparency.

\section{Conclusions}
We have measured the electron transparency of a Micromegas mesh for a range of drift fields
and have been able to reproduce these measurements with a microscopic calculation based
on Magboltz, both using finite element and using boundary element field calculations.
Zero-diffusion calculations, akin to the flux argument, fail to reproduce the measurements.
Neither do calculations which oversimplify the geometry or neglect the dipole moment of the mesh.

\acknowledgments{
The R\&D effort of the Muon ATLAS Micromegas Activity has been an important source of
inspiration for this work.
The measurements were performed in the RD51 collaboration laboratory at CERN during summer 2009.
The hospitality of the RD51 collaboration which granted access to the common
facilities is greatly appreciated.
Fruitful discussions with Dimitris Fassouliotis and Christine Kourkoumelis are acknowledged.
Supratik Mukhopadhyay and Nayana Majumdar, authors of neBEM, contributed to the calculations presented in this paper.
This work was supported in part by the U.S. Department of Energy under contract 
No.~DE-AC02-98CHI-886. 
K.N. acknowledges support from the Greek State Scholarships Foundation (I.K.Y.).}
\bibliographystyle{JHEPdoi}
\bibliography{reference.bib}
\end{document}